\documentclass[aps,prl,twocolumn,showpacs,superscriptaddress,groupedaddress]{revtex4-1}  % for review and submission

\usepackage{graphicx}
\usepackage{graphics}
\usepackage{color}
\usepackage[english]{babel}
\usepackage[T1]{fontenc}
\usepackage[latin1]{inputenc}
\usepackage{amsmath}
\usepackage{amssymb}
\usepackage{subfigure}
\usepackage{rotating}
\def\d{\mbox{d}}

\newcommand{\lra}[1]{\langle #1 \rangle }
\def     \i{\`{i}~}

\definecolor{light-red}{RGB}{255,102,102}

\begin{document}

\title{
Fluctuations of entropy production in turbulent thermal convection
}

\author{
Francesco Zonta$^{1~}$,
Sergio Chibbaro$^{2~3}$,
%...$^{4, 5~}$
}
\affiliation{
Department of Elec., Manag. and Mechanical Engineering,
University of Udine, 33100, Udine, Italy \\
$^{2}$ Sorbonne Universit\'es, UPMC Univ Paris 06, UMR 7190, Institut Jean Le Rond d'Alembert, F-75005, Paris, France\\
$^3$CNRS, UMR 7190, Institut Jean Le Rond d'Alembert, F-75005, Paris, France
}
%%%%%%%%%%%%%%%%%%%%%%%%%%%%%%%%%%%%%%%%%%%%%
%double spacing
%\dspaceon
%

%%%%%%%%%%%%%%%%%%%%% ABSTRACT %%%%%%%%%%%%%%
\begin{abstract}
 
We report on a numerical experiment performed to analyze  fluctuations of the entropy production  in turbulent thermal convection, 
  a physical configuration that represents here a prototypical case of an  out-of-equilibrium dissipative system.
We estimate the entropy production from instantaneous measurements of the local temperature and velocity fields sampled along the trajectory of a large number of point-wise Lagrangian tracers.
The entropy production is characterized by large fluctuations and becomes often negative.
This represents a sort of "finite-time" violation of the second principle of thermodynamics, 
since the direction of the energy flux  is opposite  to that prescribed by the external gradient.
We clearly show that the fluctuations of entropy production observed in the present system verify the  
fluctuation relation (FR), even though the system is time-irreversible.
\end{abstract} 

\maketitle

%%%%%%%%%%%%%%%%%%%% INTRODUCTION %%%%%%%%%%%%%%%
\emph{Introduction}
Fluctuations of physical systems close to equilibrium
are well described by the classical linear-response theory \cite{onsager1931reciprocalI,onsager1931reciprocalII,kubo1957statistical,marconi2008fluctuation}, which gives precise predictions on the behavior of the systems and leads to the fluctuation-dissipation relations.
Current knowledge of the dynamics of systems far away from equilibrium is instead much more limited.
The introduction of the so-called fluctuation relation (FR) 
~\cite{evans1993probability,PhysRevE.50.1645,gallavotti1995dynamical,gallavotti1995dynamical2,jarzynski1997nonequilibrium}
has represented a remarkable result in this area of physics. 
The FR for nonequilibrium fluctuations reduces to the Green-Kubo and Onsager relations close to equilibrium \cite{gallavotti1996extension,searles2007steady,chetrite2008fluctuation,gallavotti2014nonequilibrium} and represents one of the few exact results for systems kept far from equilibrium. However, a general response-theory for this kind of (nonequilibrium) systems  is still to be produced. 
This suggests that new analyses are required to investigate the behavior of nonequilibrium fluctuations, in particular for macroscopic 
chaotic systems \cite{ritort2004work,ciliberto2013fluctuations}.

Turbulence represents an archetype of a macroscopic dynamical system characterized by a large number of degrees of freedom and by strong fluctuations.
For its intrinsic chaotic nature, turbulence appears as a paradigmatic case to which apply, \emph{cum grano salis},
FR \cite{gallavotti2014equivalence}. 
If positively verified, this would strengthen the link between turbulence and nonequilibrium statistical mechanics. 
In particular, it would justify the hypothesis that a 
general response theory can be applied also to time-irreversible systems, at least with the purpose of computing their statistical properties.
Given the theoretical and practical importance of these issue, FR in turbulent flows has been extensively investigated in the past \cite{ciliberto1998experimental,ciliberto2004experimental,falcon2008fluctuations,cadot2008statistics,biferale1998time,gallavotti2004lyapunov,shang2005test}.
However, a satisfying statistical description of entropy fluctuations in turbulence is still to be obtained, essentially because of 
the technical problems associated to the experimental measure of fluctuations in chaotic systems \cite{zamponi2007possible} 
but also to the difficulties in performing accurate numerical simulations.
One of the central issues when discussing FR remains the choice of a convenient observable (to be measured and analyzed). 
In this work we focus on turbulent Rayleigh B\'enard convection, with the final purpose of addressing the following issues:
i) the choice of a representative observable to compute fluctuations;
ii) the presence of large deviations of this  quantity beyond the linear regime; 
iii) the applicability of FR to turbulent thermal convection.

For our scope, we use Direct Numerical Simulations and Lagrangian particle tracking of pointwise tracers, which we use as probes to measure the local thermodynamic quantities of the system.
The fundamental idea of our approach is that turbulence share similarities  with the microscopic nature of heat flows, and turbulence fluctuations correspond to thermal fluctuations \cite{ruelle2012hydrodynamic}.
With this in mind, we have chosen a configuration similar to that studied in stochastic thermodynamics \cite{sekimoto2010stochastic}, which consists of a system  kept in contact with two thermostats at different temperature and characterized by a fluctuating current. 
If our hypothesis is more than a mere analogy but a physical instance, we may expect global fluctuation relations to hold also in thermal convection. 
A key ingredient in our study is the use of a Lagrangian point of view, which 
is specifically suited to study the global transport properties of the flow \cite{toschi2009lagrangian}.
We will show that entropy production can be evaluated by looking at the work done by buoyancy on moving fluid particles. 
Provided that the correlations of the measured quantities decay fast enough,
we show that for finite time entropy production exhibits large fluctuations (being often negative) and fulfills FR.
Our results complement recent works on granular matter, a simple but complete model used to describe macroscopic irreversible systems.
For a granular matter, theory and numerical simulations agree in verifying FR, once a correct fluctuating entropy production is selected \cite{puglisi2005fluctuations,puglisi2009irreversible,sarracino2010irreversible}. 

%%%%%%%%%%%%%%%%%%%%%%%%%%%%%%%%%%%%%%%%%%
% FIG1
%%%%%%%%%%%%%%%%%%%%%%%%%%%%%%%%%%%%%%%%%%
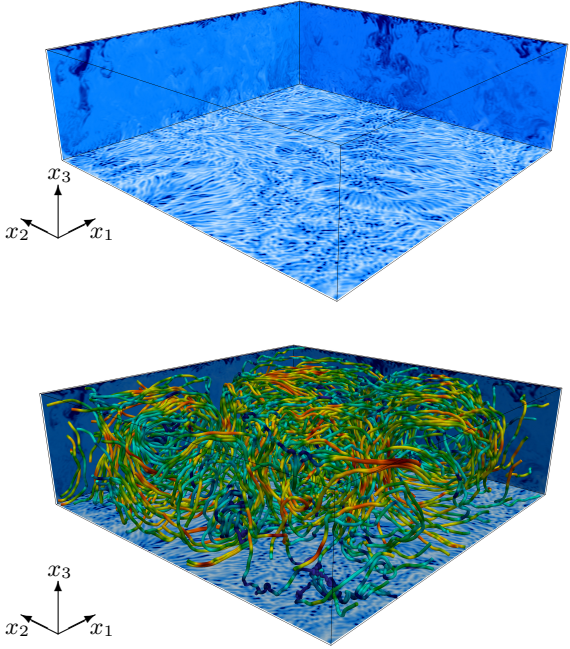
\begin{figure}[h]		  
\includegraphics[scale=.05, keepaspectratio, angle=0]{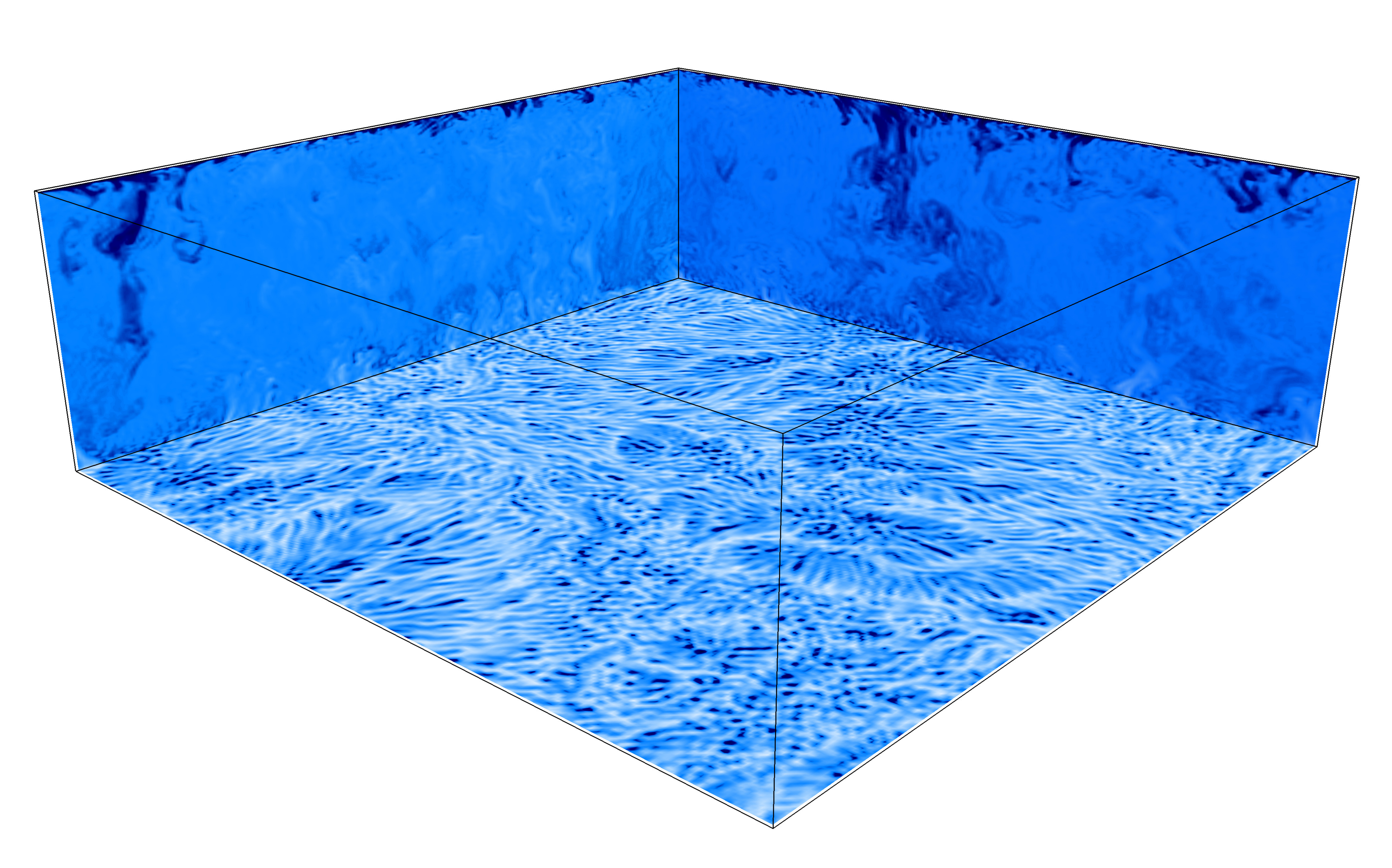}
\includegraphics[scale=.05, keepaspectratio, angle=0]{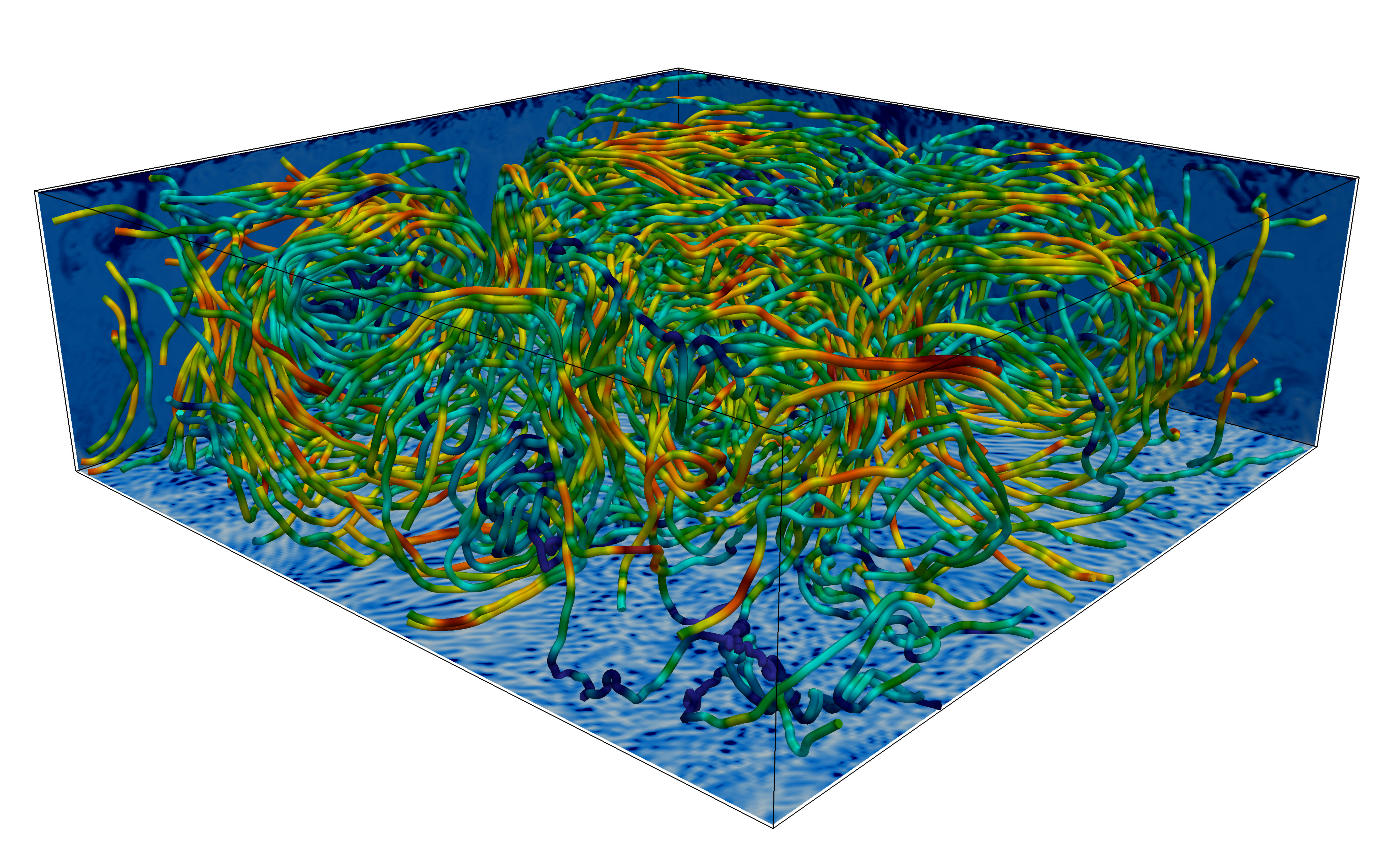}
{\hspace{-0.05cm}}
\begin{picture}(0.,0.)
\put(-200,10.){\vector(0,1){20}} 
\put(-200,10.){\vector(2,1){14}}
 \put(-200,10.){\vector(-2,1){14}}
\put(-188.,10.){$x_1$}
\put(-204.,32.){$x_3$}
\put(-220.,10.){$x_2$}
\put(-200,160.){\vector(0,1){20}} 
\put(-200,160.){\vector(2,1){14}}
 \put(-200,160.){\vector(-2,1){14}}
\put(-188.,160.){$x_1$}
\put(-204.,182.){$x_3$}
\put(-220.,160.){$x_2$}
\end{picture}

\caption{Contour maps of the temperature distribution in the convection cell for $Ra=10^9$. 
We keep the Prandtl number $Pr=4$ and we vary the Rayleigh number between $Ra=10^7$ and $Ra=10^9$.
The domain has dimensions  $L_{x_1} \times L_{x_2} \times L_{x_3}=4 \pi h \times 4 \pi h \times 2h $ and is discretized using up to $512 \times 512 \times 513$ grid nodes.
The reference velocity is the free-fall velocity $u_{ref}=(g \alpha_0 h \Delta T /2 )^{1/2}$.
(b) Example of tracers trajectories for $Ra=10^9$ colored by the local velocity magnitude.
Further details on the numerical method can be found in \cite{liot,zonta2014,zonta2013}.
}
\label{fig1}	       
\end{figure}
%%%%%%%%%%%%%%%%%%%%%%%%%%%%%%%%%%%%%%

\emph{Model}.
We consider a turbulent Rayleigh-B\'enard convection, in which an horizontal fluid layer is heated from below. 
Horizontal and wall-normal coordinates are indicated by $x_1$, $x_2$ and $x_3$, respectively.
Using the Boussinesq approximation, the system is described by the following dimensionless balance equations
\begin{eqnarray}
\frac{\partial u_i}{\partial x_i}&=&0, \\
\label{c1}
\frac{\partial u_i}{\partial t} +  u_i\frac{\partial u_i}{\partial x_j} &=& -\frac{\partial p}{\partial x_i} + 4\sqrt{\frac{Pr}{Ra}}\frac{\partial^2 u_i}{\partial x_j^2}-\delta_{i,3} \theta, \\
\label{ns1}
\frac{\partial \theta}{\partial t} +  u_i\frac{\partial \theta}{\partial x_j} &=& + \frac{4}{\sqrt{Pr Ra}}\frac{\partial^2 \theta}{\partial x_j^2}, 
\label{en1}
\end{eqnarray}
where $u_i$ is the $i^{th}$ component of the velocity vector, $p$ is pressure, $\theta= (T-T_{0})/\Delta T$ is the dimensionless temperature, $\Delta T= T_{H} - T_{C} $ is the imposed  temperature difference  between the hot bottom wall ($T_H$) and top cold wall ($T_C$), whereas $\delta_{1,3} \theta$ is the driving buoyancy force  (acting in the vertical direction $x_3$ only).
Periodicity is imposed on velocity and temperature along the horizontal directions $x_1$ and $x_2$, whereas no slip conditions are enforced for velocity at the top and bottom walls.
The fluid kinematic viscosity $\nu_{0}$, thermal diffusivity $\kappa_{0}$ and thermal expansion coefficient $\alpha_{0}$ are evaluated at the reference fluid temperature  $T_{0}=(T_H+T_C)/2\simeq 29^o C$.
The Prandtl and the Rayleigh numbers in Eqs. (\ref{c1})-(\ref{en1}) are defined as $Pr=\nu_{0}/k_{0}$ and $Ra= (g \alpha_{0} \Delta T (2h)^3) / (\nu_{0} k_{0})$, with $h=0.15~m$ the half domain height and $g$ the acceleration due to gravity. 
%%%%%%%%%%%%%%%%%%%%% RESULTS
An example of the temperature distribution inside our convection cell is given in Fig. \ref{fig1}a.
To measure the local values of the field variables, we make use of a Lagrangian approach. 
The dynamics of $N_p=1.28 \cdot 10^5$ Lagrangian tracers is computed as
\begin{equation}
\dot{\bf x}_p = {\bf u}\left( {\bf x_p}(t), t\right)~~~~~
{ \theta_p} = {\theta}\left( {\bf x_p}(t), t\right), 
\label{lag}
\end{equation} 
with ${\bf x_p}=(x_{p,1},x_{p,2},x_{p,3})$ the tracers position, $\dot{\bf x}_p=(u_{p,1},u_{p,2},u_{p,3})$ their velocity and $\theta_p$ their temperature.
A visualization of different particle trajectories (colored by the local velocity magnitude) is shown in figure \ref{fig1}b, highlighting also the chaotic nature of the flow.

Fluctuation relation (FR) concerns the symmetry of a representative observable, which is typically linked to the work done on the system, and, through dissipation, to irreversibility.
 For a Markov process, whose dynamics is described by $\dot{{\bf x}}(t)={\bf u}({\bf x}(t),t)$, the representative observable can be written as~\cite{lebowitz1982steady}
\begin{equation}
\beta W_t= \frac{1}{t}\log \frac{\Pi(\{{\bf x}(s)\}_0^t)}{\Pi(\{I{\bf x}(s)\}_0^t)}~,
\end{equation}
where $\{{\bf x}(s)\}_0^t$ and $\{I{\bf x}(s)\}_0^t$ are  the direct and the time-reversed trajectories in
 the time interval $[0, t]$ respectively, while $\Pi$ indicates probability and $\beta$ is a suitable 
energy scale of the system (for small systems $\beta^{-1}=k_B T$, with $k_B$ the Boltzmann constant and $T$ the absolute temperature of the system).
When FR applies, 
\begin{equation}
\log \frac{\Pi(\beta W_t=p)}{\Pi(\beta W_t=-p)}=t p,
\end{equation}
where $W_t$ is usually called "produced entropy".
Note that the choice of a representative observable quantity is  crucial  to verify the fluctuation relation. 
Entropy production for a Markov process is usually defined based on the dynamical probability,
a quantity that is generally not accessible in complex systems.
Previous experimental studies on macroscopic chaotic systems focused on the behavior of the injected power, a measurable quantity that however 
was found to depart from the predictions given by the  FR \cite{farago2002injected,puglisi2005fluctuations}.
In the present case, we start from the balance equation for the turbulent kinetic energy $E_t=1/2  \lra{u'_i u'_i}$, where brackets $\langle \rangle$ indicate statistical average  and $u'_i=u_i-\langle u_i\rangle$  velocity fluctuations.
Since the system is homogeneous along the $x_1$ and $x_2$ directions, we obtain 
\begin{eqnarray}
\frac{\partial E_t}{\partial t}+\frac{\partial  }{\partial x_3}(\langle E_t u'_3 \rangle +  \lra{u'_3 p'})    = 
 \langle u_3' \theta' \rangle - \langle \epsilon \rangle ,
\label{budget}
\end{eqnarray}
where $\langle \epsilon \rangle =4\sqrt{\frac{Pr}{Ra}}\left(\frac{ \partial^2 \langle E_t \rangle}{\partial x^2_3}
+ \left \langle \frac{ \partial u'_i}{\partial x_j}\frac{\partial u'_i}{\partial x_j} \right \rangle \right)$ is the turbulent dissipation and $\theta'$ is  the temperature fluctuation.
The volume-averaged steady state solution gives  $ \langle u_3' \theta' \rangle = \langle \epsilon \rangle$. 
This provides also an estimate of the entropy production, which from thermodynamics is   $\lra{\epsilon} / \lra{T}$. 
Specifically, we focus on the term $W= \rho_{0} g \alpha_0  \theta_p' u'_{3,p}  $, which we measure along the path of the Lagrangian tracers  and 
 that represents the power spent by buoyancy to displace a fluid parcel. 
Note that $W$ quantifies the vertical flux of energy and is therefore also linked to the vertical Nusselt number 
\cite{gasteuil,liot}.
Warm fluctuations  $\theta_p'>0$ produce a positive energy flux 
when associated to positive vertical velocities $u_{p,3}>0$,
whereas cold fluctuations $\theta_p'<0$ produce a positive energy flux when associated to negative vertical velocity $u_{p,3}<0$.

We now consider the time-averaged (but fluctuating)  expression of the global energy balance of the system, 
$\Delta U_{\tau}=W_{\tau}+Q_{\tau}$, where
\begin{equation}
W_{\tau}=\frac{1}{\tau}\int_0^{\tau} {\bf F}^{ext}(t) \cdot {\bf u}(t) \mathrm{dt} = \frac{1}{\tau}\int_0^{\tau} \rho_{0} g \alpha_0 \theta_p^{\prime}u^{\prime}_{p,3} \mathrm{dt}
\end{equation}
and ${\bf F^{ext}}$ is the external force field  due to gravity, whereas $\Delta U_{\tau}$ and $Q_{\tau}$ indicate internal energy and heat, respectively.
In the limit ${\tau} \to \infty$, $\lra{\Delta U_{\tau}}= 0$ and the energy balance of the 
system becomes $\langle W_{\tau} \rangle=\langle Q_{\tau} \rangle$.
Although work and heat fluctuations are generally different in stochastic systems  \cite{van2004extended}, 
we believe that  $W_{\tau}$ can give a good estimate of entropy production.

\emph{Results}
%%%%%%%%%%%%%%%%%%%%%%%%%%%%%%%%%%%%%%%%%%
\vskip -2.5cm
\begin{figure}[h]		  
\includegraphics[scale=.35, keepaspectratio, angle=0]{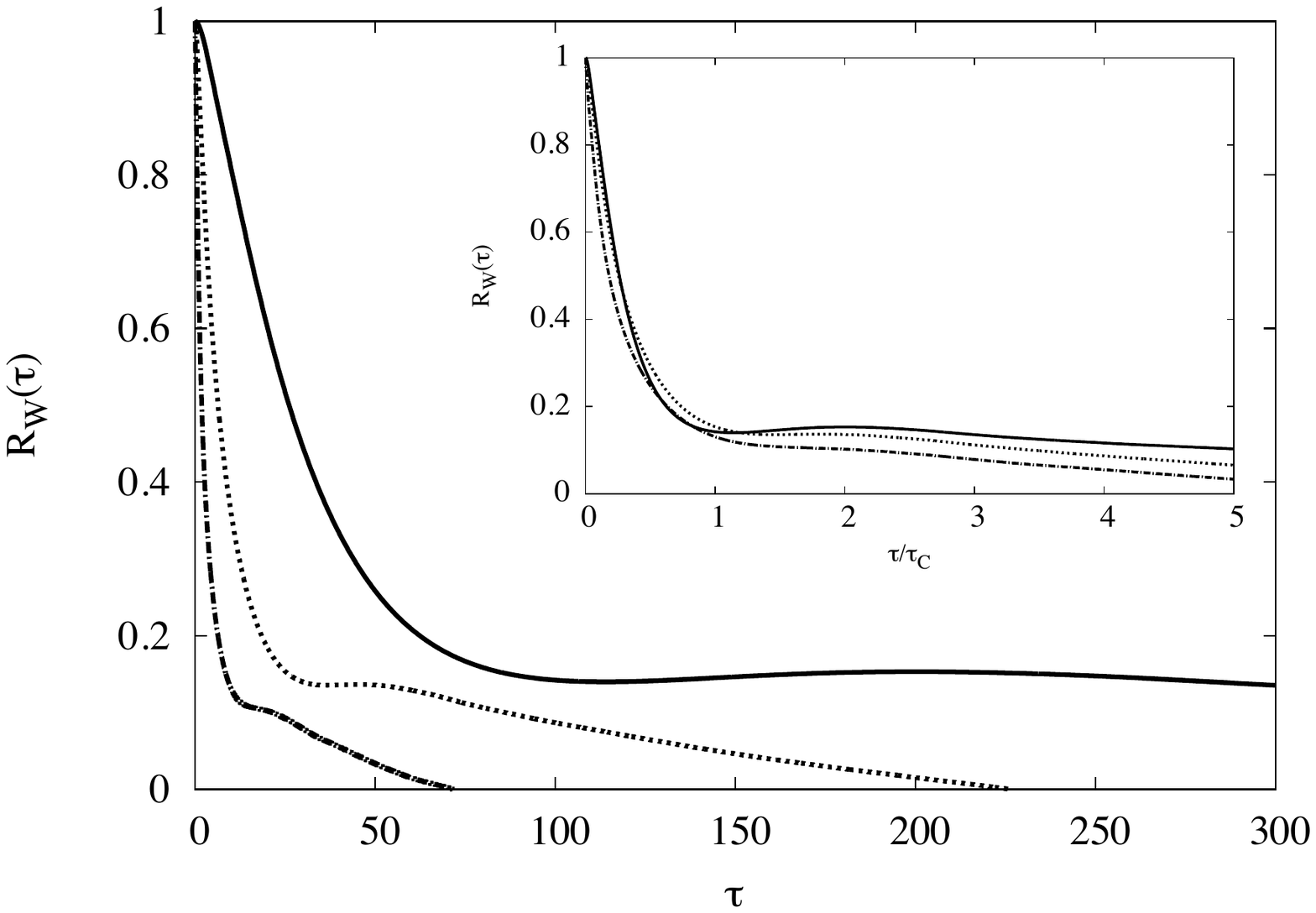}
%\setlength{\unitlength}{3.7cm}
%\begin{picture}(0.,0.)
%%\put(-.41,+.62){\vector(-1,-1){0.33}} 
%%\put(-.4,+.64){$Ra$}
%\end{picture}
\vskip -2.5cm
\caption{ Lagrangian correlation $R_{W}$ as a function of the time lag $\tau$ for the three different Rayleigh number.
The collapse of the correlation function upon rescaling of the time lag $\tau$ by the integral time scale $\tau_C$ is explicitly shown in the inset.
Note that 
$\tau_C=100~s$ for $Ra=10^7$, 
$\tau_C=25~s$ for $Ra=10^8$ and
$\tau_C=7~s$ for $Ra=10^9$.
}
\label{corr}	       
\end{figure}
%%%%%%%%%%%%%%%%%%%%%%%%%%%%%%%%%%%%%%
In the following, we will assume that time-averages are equivalent to ensemble averages (ergodicity).
This assumption is justified if the correlation function of $W$, $R_W=\langle W(t)W(t+\tau)\rangle/\langle W^2\rangle$, computed after a statistically steady state is reached, exhibits a fast decrease in time \cite{monin2007statistical}.
%%%%
To verify this, we explicitly compute $R_W(\tau)$ for each $Ra$ as a function of the time lag $\tau$.
Results are shown in Fig. \ref{corr}.
We observe that $R_W$ is characterized by an exponential decay $\exp (-t/\Gamma)$, whose decay rate $1/\Gamma$ increases with increasing $Ra$.
As a consequence, velocity and temperature fluctuations decorrelate faster for  large $Ra$, due to the larger fluctuations observed for increasing $Ra$.
The fast decay of the correlation function happens also in Anosov dynamical systems and in Markov processes, and is a key feature  
to obtain steady FR\cite{marconi2008fluctuation}.
From the behavior of the correlation function $R_{W}$, we are able to compute the integral correlation time $\tau_C=\int_{0}^{\infty}R_{W}(\tau) \d \tau$.
Upon rescaling of the time lag $\tau$ by $\tau_C$, the correlation functions $R_W$ computed at different $Ra$ collapse (inset of Fig. \ref{corr}).
For $\tau/\tau_C>1$ the value of the correlation is already $R_W<0.2$, indicating that from $\tau\simeq \tau_C$ the signal is only barely reminiscent of the starting condition. This means that $ \tau_C$ is a representative time scale of the system and suggests that FR can be conveniently tested for $ \tau/\tau_C>1$.
%%%%%%%%%%%%%%%%%%%%%%%%%%%%%%%%%%%%%%%%%%
\vskip -2.5cm
\begin{figure}[h]		  
\includegraphics[scale=.35]{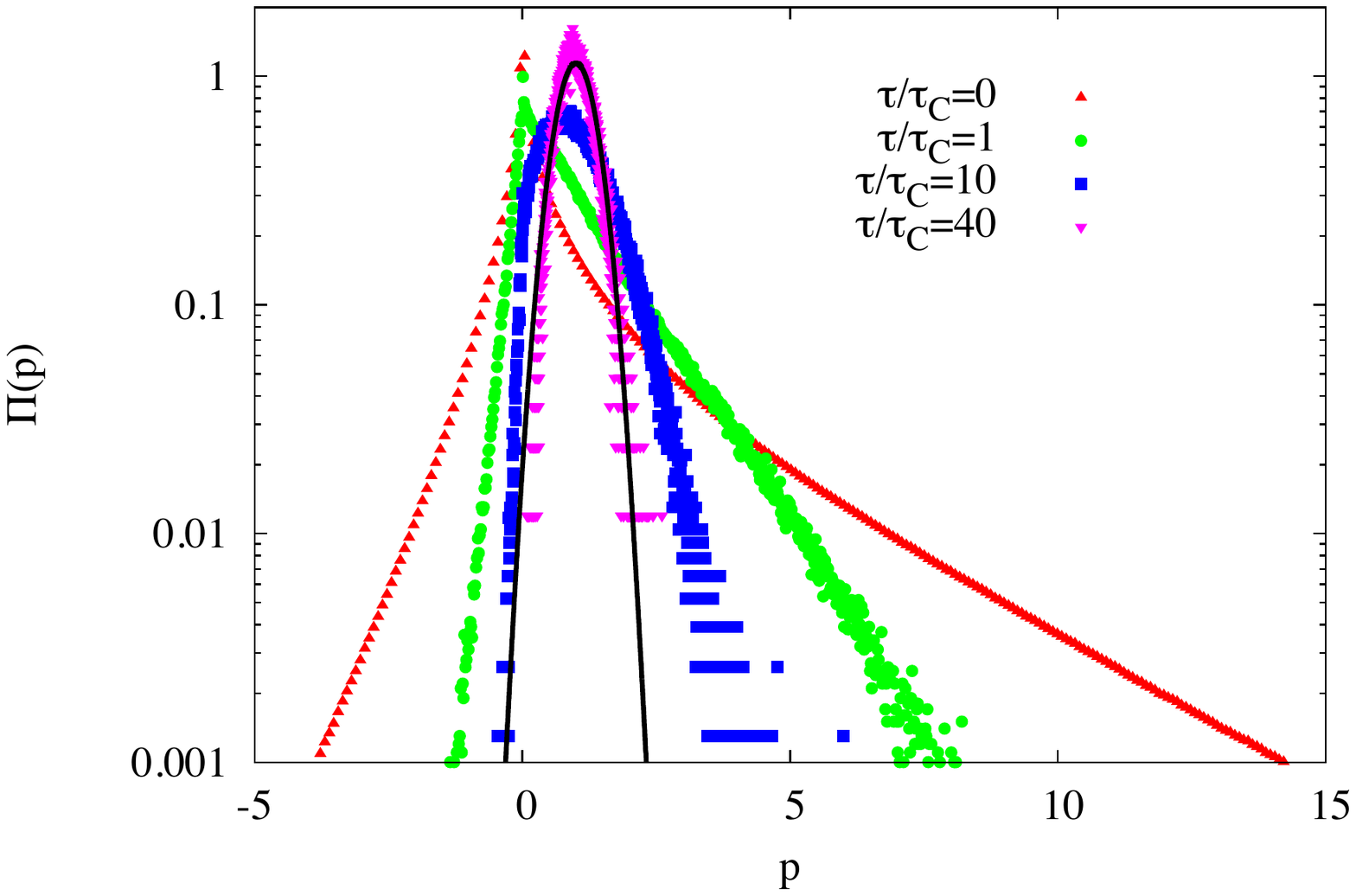}
\vskip -2.5cm
\caption{Probability density function ($\Pi$) of the normalized energy flux 
$p=W_{\tau}/\langle W_{\tau} \rangle$ for $Ra = 10^9$ and for different values of the averaging time window $\tau/\tau_C$. 
The behavior of a Gaussian distribution is explicitly indicated by the solid line.
Results from simulations at $Ra = 10^7$ and  $Ra = 10^8$ are not shown here because they are qualitatively 
similar to those at $Ra=10^9$ and do not add to the discussion.
}
\label{pdf}	       
\end{figure}
%%%%%%%%%%%%%%%%%%%%%%%%%%%%%%%%%%%%%%

Starting from the Lagrangian measurements of $W$, we compute the probability density  function ($\Pi$) of the normalized functional  $p=\frac{W_{\tau}}{\lra{W_{\tau}}}$ for different values of $\tau$,
 shown in Fig. \ref{pdf}.
For $\tau/\tau_C=0$, 
$\Pi(p)$ is highly asymmetric, with the most probable value occurring for $p=0$ and with positive fluctuations being larger than negative ones.
The asymmetry of $\Pi(p)$ persists also for increasing $\tau/\tau_C$ and disappears only when the average is done on a rather large time window ($\tau/\tau_C \ge 40$).
In particular, for $\tau/\tau_C \simeq 40$ the distribution peaks around $p\simeq 1$ and recovers an almost gaussian distribution.
Note that at this stage ($\tau/\tau_C \ge 40$), the probability of negative events becomes essentially zero.
These observations suggest that, although the imposed mean temperature difference between the walls induces a net positive vertical energy flux ( $\langle W \rangle > 0$), $W$ can often be negative.
The occurrence of countegradient fluxes of global transport properties (such as the Nusselt number) is an extremely important phenomenon that has been also observed in other situations \cite{lohse2}.
From a physical point of view, small positive and negative values of $W$ are produced by turbulence, which is uncorrelated with the temperature field.
These small positive and negative values of $W$ balance each others and do not contribute to the average heat transport \cite{shang2005test}. 
Only large velocity and temperature fluctuations produced by thermal plumes (rising hot plumes and falling cold plumes) are correlated and contribute to the positive mean heat flux.

\begin{figure}[h]		  
\includegraphics[scale=.35]{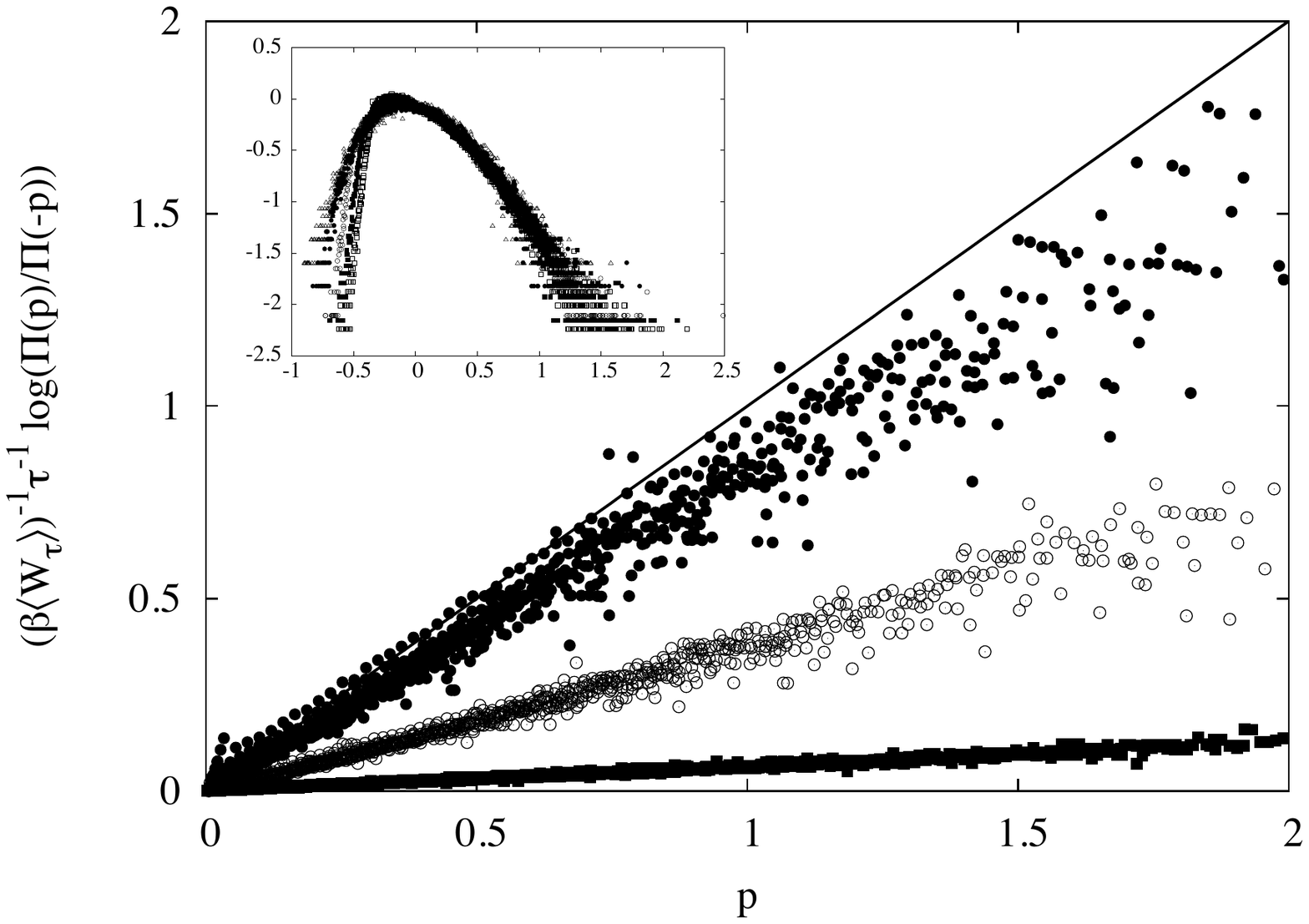}
%\setlength{\unitlength}{3.7cm}
%\begin{picture}(0.,0.)
%\put(+.81,+.25){\vector(-1,2){0.5}} 
%\put(+.2,+1.3){$Ra$}
%\end{picture}
%\vskip -2.5cm
\caption{Behavior of $(\beta \lra{W_{\tau}})^{-1}\tau^{-1}\log \left( \Pi(p)/\Pi(-p) \right)$  as a function of $p$ for $1<\tau/\tau_C<30$, i.e. when the probability density function is not gaussian. Results are shown for each value of the Rayleigh number $Ra$ (as indicated by the arrow).
The solid line represents the theoretical prediction given by the FR, a linear function with slope equal to unity.
In the inset, the Cram\'er function $\zeta(p)$, rescaled as suggested by ~\cite{rondoni2003large} using its standard deviation $\sigma_{\tau}$ and its slope $C_{\tau}$, is shown for different $\tau$ (for $1<\tau<30$). The convergence is satisfying for $\tau>1$. 
}
\label{slope}	       
\end{figure}

\begin{figure}[h]		  
\includegraphics[scale=.35]{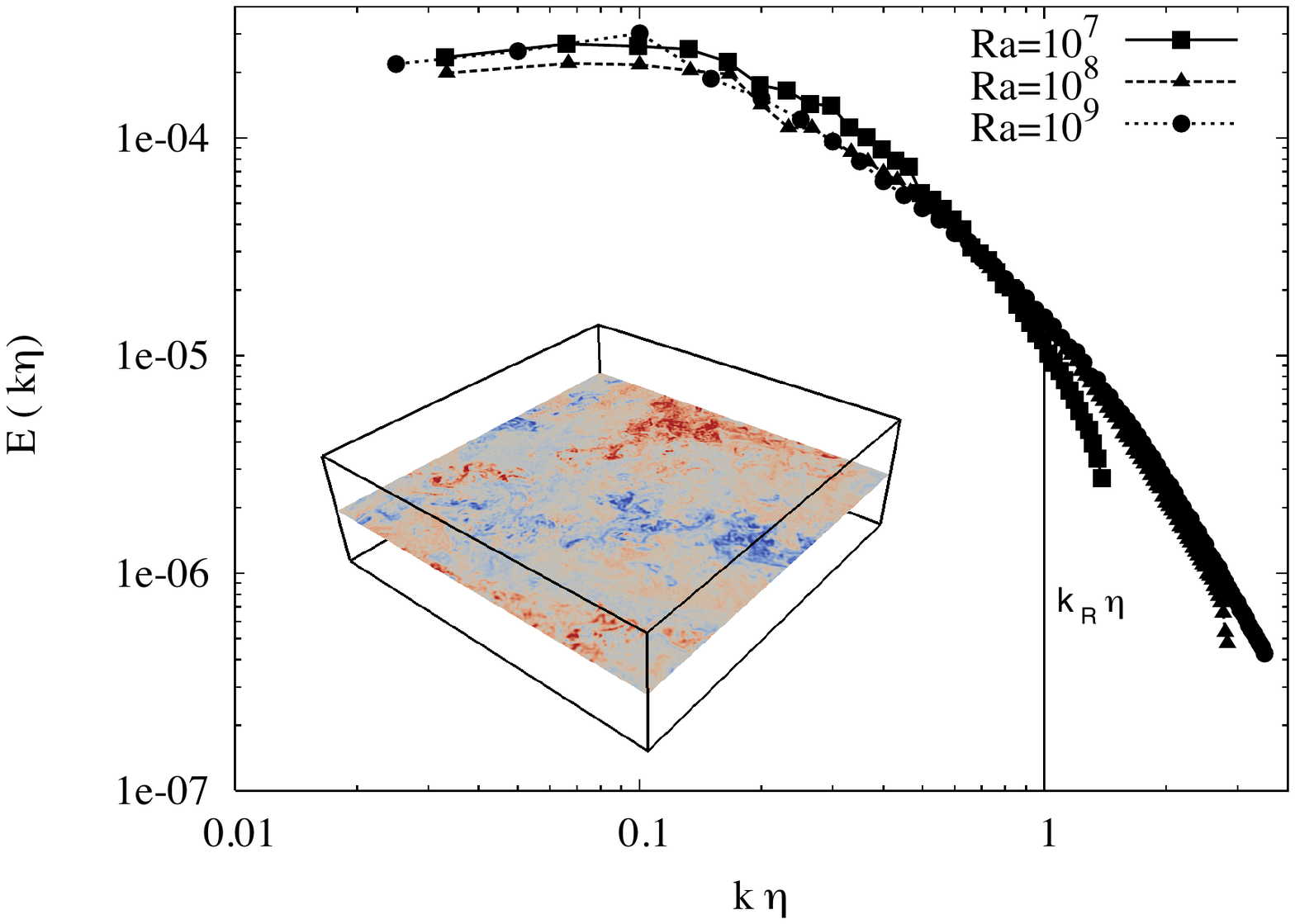}
\caption{Time-averaged energy spectra of the turbulent kinetic energy computed at the center of the channel (on the central plane explicitly shown in the inset of the figure, where the flow appears to be homogeneous). 
The representative energy scale of the system $\beta^{-1}=(k_B T)_{turb}=1/2 \rho \int_{k>k_R}E(k) \d k$, computed assuming $k_R \eta=1$,  
gives $\beta=0.49$ for $Ra=10^7$, $\beta=0.33$ for $Ra=10^8$ and $\beta=0.3$ for $Ra=10^9$.
}
\label{spectra}	       
\end{figure}

It is reasonable to expect that the fluctuations of $p$ are governed by a \emph{large deviation} law $\Pi(p)\sim e^{\tau\zeta(p)}$, with $\zeta$ concave.
Then, from the behavior of $\Pi(p)$, we measure the quantity 
\begin{equation}
\zeta(p)-\zeta(-p)=\sigma(p) = \frac{1} {\tau} \log \frac{\Pi(p)}{\Pi(-p)}
\end{equation}
for different averaging time $\tau/\tau_C$ taken in the range $1<\tau/\tau_C<30$.
The resulting  behavior, given in  
 Fig. \ref{slope}, nicely shows that $\sigma(p)$ is a linear function of $p$,
\begin{equation}
\sigma(p) = \alpha p .
\end{equation}
In particular, we observe that  the slope $\alpha$ of the curve increases with increasing $Ra$ and tends  (for $Ra=10^9$) to $\alpha=\beta \lra{W_{\tau}}$,  as theoretically predicted by the FR. 
As already mentioned, $\beta$ is a suitable and representative energy scale of the system.
Following the  Kolmogorov cascade picture~\cite{gallavotti1997dynamical,rondoni1999fluctuations}, to study the microscopic nature of the fluctuations we assume that $\beta^{-1}$ is the energy of the dissipative scales 
$\beta^{-1}=(k_B T)_{turb}=1/2 \rho \int_{k>k_R}E(k) \d k$, with $k_R$ the wavenumber characterizing dissipation (\emph{ i.e.} $K_R \eta \simeq 1$, with $\eta$ the Kolmogorov lengthscale).
The value of $\beta$ is obtained from the turbulent kinetic energy spectrum $E(k)$, as shown in Fig. \ref{spectra}.
%
%%%%%%%%%%%%%%%%%%%%%%%%%%%%%%%%%%%%%%%%%%
%\vskip -2.5cm
%%%%%%%%%%%%%%%%%%%%%%%%%%%%%%%%%%%%%%

\emph{Conclusion} 
In this letter, we have used a Lagrangian approach to study fluctuations of entropy production in turbulent thermal convection.
Entropy production has been evaluated from the local measurements of the quantity $W_{\tau} \propto   \theta'_p u'_{3,p}$ (work done by buoyancy on fluid particles) along the trajectory of individual Lagrangian tracers. 
We have shown that $W_{\tau}$ is often negative, and is characterized by fluctuations that follow the FR beyond the linear regime, provided that a representative energy scale for 
dissipation is suitably identified. 
This result sheds new light on turbulence, allowing an \emph{a priori} estimate of the behavior of fluctuations of energy flux or entropy production  and giving access 
to the Cram\'er function.
%%%%%%%%%%%%%%%%%%%%%%%%%%%%%%%%%%%%%%%%%%
%%%%%%%%%%%%%%%%%%%%%%%%%%%%%%%%%%%%%%
\\
We acknowledge Massimo Cencini, Andrea Crisanti 
%Andrea Puglisi, Lamberto Rondoni, 
and Dario Villamaina 
%and Angelo Vulpiani 
 for fruitful discussions.
%%%%%%%%%%%%%%%%%%%%%% REFERENCES %%%%%%%%%%%%%%%%
\bibliography{biblio-zs}

\begin{thebibliography}{42}%
\makeatletter
\providecommand \@ifxundefined [1]{%
 \@ifx{#1\undefined}
}%
\providecommand \@ifnum [1]{%
 \ifnum #1\expandafter \@firstoftwo
 \else \expandafter \@secondoftwo
 \fi
}%
\providecommand \@ifx [1]{%
 \ifx #1\expandafter \@firstoftwo
 \else \expandafter \@secondoftwo
 \fi
}%
\providecommand \natexlab [1]{#1}%
\providecommand \enquote  [1]{``#1''}%
\providecommand \bibnamefont  [1]{#1}%
\providecommand \bibfnamefont [1]{#1}%
\providecommand \citenamefont [1]{#1}%
\providecommand \href@noop [0]{\@secondoftwo}%
\providecommand \href [0]{\begingroup \@sanitize@url \@href}%
\providecommand \@href[1]{\@@startlink{#1}\@@href}%
\providecommand \@@href[1]{\endgroup#1\@@endlink}%
\providecommand \@sanitize@url [0]{\catcode `\\12\catcode `\$12\catcode
  `\&12\catcode `\#12\catcode `\^12\catcode `\_12\catcode `\%12\relax}%
\providecommand \@@startlink[1]{}%
\providecommand \@@endlink[0]{}%
\providecommand \url  [0]{\begingroup\@sanitize@url \@url }%
\providecommand \@url [1]{\endgroup\@href {#1}{\urlprefix }}%
\providecommand \urlprefix  [0]{URL }%
\providecommand \Eprint [0]{\href }%
\providecommand \doibase [0]{http://dx.doi.org/}%
\providecommand \selectlanguage [0]{\@gobble}%
\providecommand \bibinfo  [0]{\@secondoftwo}%
\providecommand \bibfield  [0]{\@secondoftwo}%
\providecommand \translation [1]{[#1]}%
\providecommand \BibitemOpen [0]{}%
\providecommand \bibitemStop [0]{}%
\providecommand \bibitemNoStop [0]{.\EOS\space}%
\providecommand \EOS [0]{\spacefactor3000\relax}%
\providecommand \BibitemShut  [1]{\csname bibitem#1\endcsname}%
\let\auto@bib@innerbib\@empty
%</preamble>
\bibitem [{\citenamefont
  {Onsager}(1931{\natexlab{a}})}]{onsager1931reciprocalI}%
  \BibitemOpen
  \bibfield  {author} {\bibinfo {author} {\bibfnamefont {L.}~\bibnamefont
  {Onsager}},\ }\href@noop {} {\bibfield  {journal} {\bibinfo  {journal}
  {Physical Review}\ }\textbf {\bibinfo {volume} {37}},\ \bibinfo {pages} {405}
  (\bibinfo {year} {1931}{\natexlab{a}})}\BibitemShut {NoStop}%
\bibitem [{\citenamefont
  {Onsager}(1931{\natexlab{b}})}]{onsager1931reciprocalII}%
  \BibitemOpen
  \bibfield  {author} {\bibinfo {author} {\bibfnamefont {L.}~\bibnamefont
  {Onsager}},\ }\href@noop {} {\bibfield  {journal} {\bibinfo  {journal}
  {Physical Review}\ }\textbf {\bibinfo {volume} {38}},\ \bibinfo {pages}
  {2265} (\bibinfo {year} {1931}{\natexlab{b}})}\BibitemShut {NoStop}%
\bibitem [{\citenamefont {Kubo}(1957)}]{kubo1957statistical}%
  \BibitemOpen
  \bibfield  {author} {\bibinfo {author} {\bibfnamefont {R.}~\bibnamefont
  {Kubo}},\ }\href@noop {} {\bibfield  {journal} {\bibinfo  {journal} {Journal
  of the Physical Society of Japan}\ }\textbf {\bibinfo {volume} {12}},\
  \bibinfo {pages} {570} (\bibinfo {year} {1957})}\BibitemShut {NoStop}%
\bibitem [{\citenamefont {Marconi}\ \emph {et~al.}(2008)\citenamefont
  {Marconi}, \citenamefont {Puglisi}, \citenamefont {Rondoni},\ and\
  \citenamefont {Vulpiani}}]{marconi2008fluctuation}%
  \BibitemOpen
  \bibfield  {author} {\bibinfo {author} {\bibfnamefont {U.~M.~B.}\
  \bibnamefont {Marconi}}, \bibinfo {author} {\bibfnamefont {A.}~\bibnamefont
  {Puglisi}}, \bibinfo {author} {\bibfnamefont {L.}~\bibnamefont {Rondoni}}, \
  and\ \bibinfo {author} {\bibfnamefont {A.}~\bibnamefont {Vulpiani}},\
  }\href@noop {} {\bibfield  {journal} {\bibinfo  {journal} {Physics reports}\
  }\textbf {\bibinfo {volume} {461}},\ \bibinfo {pages} {111} (\bibinfo {year}
  {2008})}\BibitemShut {NoStop}%
\bibitem [{\citenamefont {Evans}\ \emph {et~al.}(1993)\citenamefont {Evans},
  \citenamefont {Cohen},\ and\ \citenamefont {Morriss}}]{evans1993probability}%
  \BibitemOpen
  \bibfield  {author} {\bibinfo {author} {\bibfnamefont {D.~J.}\ \bibnamefont
  {Evans}}, \bibinfo {author} {\bibfnamefont {E.}~\bibnamefont {Cohen}}, \ and\
  \bibinfo {author} {\bibfnamefont {G.}~\bibnamefont {Morriss}},\ }\href@noop
  {} {\bibfield  {journal} {\bibinfo  {journal} {Physical Review Letters}\
  }\textbf {\bibinfo {volume} {71}},\ \bibinfo {pages} {2401} (\bibinfo {year}
  {1993})}\BibitemShut {NoStop}%
\bibitem [{\citenamefont {Evans}\ and\ \citenamefont
  {Searles}(1994)}]{PhysRevE.50.1645}%
  \BibitemOpen
  \bibfield  {author} {\bibinfo {author} {\bibfnamefont {D.~J.}\ \bibnamefont
  {Evans}}\ and\ \bibinfo {author} {\bibfnamefont {D.~J.}\ \bibnamefont
  {Searles}},\ }\href {\doibase 10.1103/PhysRevE.50.1645} {\bibfield  {journal}
  {\bibinfo  {journal} {Phys. Rev. E}\ }\textbf {\bibinfo {volume} {50}},\
  \bibinfo {pages} {1645} (\bibinfo {year} {1994})}\BibitemShut {NoStop}%
\bibitem [{\citenamefont {Gallavotti}\ and\ \citenamefont
  {Cohen}(1995{\natexlab{a}})}]{gallavotti1995dynamical}%
  \BibitemOpen
  \bibfield  {author} {\bibinfo {author} {\bibfnamefont {G.}~\bibnamefont
  {Gallavotti}}\ and\ \bibinfo {author} {\bibfnamefont {E.}~\bibnamefont
  {Cohen}},\ }\href@noop {} {\bibfield  {journal} {\bibinfo  {journal}
  {Physical Review Letters}\ }\textbf {\bibinfo {volume} {74}},\ \bibinfo
  {pages} {2694} (\bibinfo {year} {1995}{\natexlab{a}})}\BibitemShut {NoStop}%
\bibitem [{\citenamefont {Gallavotti}\ and\ \citenamefont
  {Cohen}(1995{\natexlab{b}})}]{gallavotti1995dynamical2}%
  \BibitemOpen
  \bibfield  {author} {\bibinfo {author} {\bibfnamefont {G.}~\bibnamefont
  {Gallavotti}}\ and\ \bibinfo {author} {\bibfnamefont {E.}~\bibnamefont
  {Cohen}},\ }\href@noop {} {\bibfield  {journal} {\bibinfo  {journal} {Journal
  of Statistical Physics}\ }\textbf {\bibinfo {volume} {80}},\ \bibinfo {pages}
  {931} (\bibinfo {year} {1995}{\natexlab{b}})}\BibitemShut {NoStop}%
\bibitem [{\citenamefont {Jarzynski}(1997)}]{jarzynski1997nonequilibrium}%
  \BibitemOpen
  \bibfield  {author} {\bibinfo {author} {\bibfnamefont {C.}~\bibnamefont
  {Jarzynski}},\ }\href@noop {} {\bibfield  {journal} {\bibinfo  {journal}
  {Physical Review Letters}\ }\textbf {\bibinfo {volume} {78}},\ \bibinfo
  {pages} {2690} (\bibinfo {year} {1997})}\BibitemShut {NoStop}%
\bibitem [{\citenamefont {Gallavotti}(1996)}]{gallavotti1996extension}%
  \BibitemOpen
  \bibfield  {author} {\bibinfo {author} {\bibfnamefont {G.}~\bibnamefont
  {Gallavotti}},\ }\href@noop {} {\bibfield  {journal} {\bibinfo  {journal}
  {Physical Review Letters}\ }\textbf {\bibinfo {volume} {77}},\ \bibinfo
  {pages} {4334} (\bibinfo {year} {1996})}\BibitemShut {NoStop}%
\bibitem [{\citenamefont {Searles}\ \emph {et~al.}(2007)\citenamefont
  {Searles}, \citenamefont {Rondoni},\ and\ \citenamefont
  {Evans}}]{searles2007steady}%
  \BibitemOpen
  \bibfield  {author} {\bibinfo {author} {\bibfnamefont {D.~J.}\ \bibnamefont
  {Searles}}, \bibinfo {author} {\bibfnamefont {L.}~\bibnamefont {Rondoni}}, \
  and\ \bibinfo {author} {\bibfnamefont {D.~J.}\ \bibnamefont {Evans}},\
  }\href@noop {} {\bibfield  {journal} {\bibinfo  {journal} {Journal of
  Statistical Physics}\ }\textbf {\bibinfo {volume} {128}},\ \bibinfo {pages}
  {1337} (\bibinfo {year} {2007})}\BibitemShut {NoStop}%
\bibitem [{\citenamefont {Chetrite}\ and\ \citenamefont
  {Gawedzki}(2008)}]{chetrite2008fluctuation}%
  \BibitemOpen
  \bibfield  {author} {\bibinfo {author} {\bibfnamefont {R.}~\bibnamefont
  {Chetrite}}\ and\ \bibinfo {author} {\bibfnamefont {K.}~\bibnamefont
  {Gawedzki}},\ }\href@noop {} {\bibfield  {journal} {\bibinfo  {journal}
  {Communications in Mathematical Physics}\ }\textbf {\bibinfo {volume}
  {282}},\ \bibinfo {pages} {469} (\bibinfo {year} {2008})}\BibitemShut
  {NoStop}%
\bibitem [{\citenamefont {Gallavotti}(2014)}]{gallavotti2014nonequilibrium}%
  \BibitemOpen
  \bibfield  {author} {\bibinfo {author} {\bibfnamefont {G.}~\bibnamefont
  {Gallavotti}},\ }\href@noop {} {\emph {\bibinfo {title} {Nonequilibrium and
  irreversibility}}}\ (\bibinfo  {publisher} {Springer},\ \bibinfo {year}
  {2014})\BibitemShut {NoStop}%
\bibitem [{\citenamefont {Ritort}(2004)}]{ritort2004work}%
  \BibitemOpen
  \bibfield  {author} {\bibinfo {author} {\bibfnamefont {F.}~\bibnamefont
  {Ritort}},\ }in\ \href@noop {} {\emph {\bibinfo {booktitle} {Poincar{\'e}
  Seminar 2003}}}\ (\bibinfo {organization} {Springer},\ \bibinfo {year}
  {2004})\ pp.\ \bibinfo {pages} {193--226}\BibitemShut {NoStop}%
\bibitem [{\citenamefont {Ciliberto}\ \emph {et~al.}(2013)\citenamefont
  {Ciliberto}, \citenamefont {Gomez-Solano},\ and\ \citenamefont
  {Petrosyan}}]{ciliberto2013fluctuations}%
  \BibitemOpen
  \bibfield  {author} {\bibinfo {author} {\bibfnamefont {S.}~\bibnamefont
  {Ciliberto}}, \bibinfo {author} {\bibfnamefont {R.}~\bibnamefont
  {Gomez-Solano}}, \ and\ \bibinfo {author} {\bibfnamefont {A.}~\bibnamefont
  {Petrosyan}},\ }\href@noop {} {\bibfield  {journal} {\bibinfo  {journal}
  {Annu. Rev. Condens. Matter Phys.}\ }\textbf {\bibinfo {volume} {4}},\
  \bibinfo {pages} {235} (\bibinfo {year} {2013})}\BibitemShut {NoStop}%
\bibitem [{\citenamefont {Gallavotti}\ and\ \citenamefont
  {Lucarini}(2014)}]{gallavotti2014equivalence}%
  \BibitemOpen
  \bibfield  {author} {\bibinfo {author} {\bibfnamefont {G.}~\bibnamefont
  {Gallavotti}}\ and\ \bibinfo {author} {\bibfnamefont {V.}~\bibnamefont
  {Lucarini}},\ }\href@noop {} {\bibfield  {journal} {\bibinfo  {journal}
  {Journal of Statistical Physics}\ }\textbf {\bibinfo {volume} {156}},\
  \bibinfo {pages} {1027} (\bibinfo {year} {2014})}\BibitemShut {NoStop}%
\bibitem [{\citenamefont {Ciliberto}\ and\ \citenamefont
  {Laroche}(1998)}]{ciliberto1998experimental}%
  \BibitemOpen
  \bibfield  {author} {\bibinfo {author} {\bibfnamefont {S.}~\bibnamefont
  {Ciliberto}}\ and\ \bibinfo {author} {\bibfnamefont {C.}~\bibnamefont
  {Laroche}},\ }\href@noop {} {\bibfield  {journal} {\bibinfo  {journal} {Le
  Journal de Physique IV}\ }\textbf {\bibinfo {volume} {8}},\ \bibinfo {pages}
  {Pr6} (\bibinfo {year} {1998})}\BibitemShut {NoStop}%
\bibitem [{\citenamefont {Ciliberto}\ \emph {et~al.}(2004)\citenamefont
  {Ciliberto}, \citenamefont {Garnier}, \citenamefont {Hernandez},
  \citenamefont {Lacpatia}, \citenamefont {Pinton},\ and\ \citenamefont
  {Chavarria}}]{ciliberto2004experimental}%
  \BibitemOpen
  \bibfield  {author} {\bibinfo {author} {\bibfnamefont {S.}~\bibnamefont
  {Ciliberto}}, \bibinfo {author} {\bibfnamefont {N.}~\bibnamefont {Garnier}},
  \bibinfo {author} {\bibfnamefont {S.}~\bibnamefont {Hernandez}}, \bibinfo
  {author} {\bibfnamefont {C.}~\bibnamefont {Lacpatia}}, \bibinfo {author}
  {\bibfnamefont {J.-F.}\ \bibnamefont {Pinton}}, \ and\ \bibinfo {author}
  {\bibfnamefont {G.~R.}\ \bibnamefont {Chavarria}},\ }\href@noop {} {\bibfield
   {journal} {\bibinfo  {journal} {Physica A: Statistical Mechanics and its
  Applications}\ }\textbf {\bibinfo {volume} {340}},\ \bibinfo {pages} {240}
  (\bibinfo {year} {2004})}\BibitemShut {NoStop}%
\bibitem [{\citenamefont {Falcon}\ \emph {et~al.}(2008)\citenamefont {Falcon},
  \citenamefont {Auma{\^\i}tre}, \citenamefont {Falc{\'o}n}, \citenamefont
  {Laroche},\ and\ \citenamefont {Fauve}}]{falcon2008fluctuations}%
  \BibitemOpen
  \bibfield  {author} {\bibinfo {author} {\bibfnamefont {E.}~\bibnamefont
  {Falcon}}, \bibinfo {author} {\bibfnamefont {S.}~\bibnamefont
  {Auma{\^\i}tre}}, \bibinfo {author} {\bibfnamefont {C.}~\bibnamefont
  {Falc{\'o}n}}, \bibinfo {author} {\bibfnamefont {C.}~\bibnamefont {Laroche}},
  \ and\ \bibinfo {author} {\bibfnamefont {S.}~\bibnamefont {Fauve}},\
  }\href@noop {} {\bibfield  {journal} {\bibinfo  {journal} {Physical review
  letters}\ }\textbf {\bibinfo {volume} {100}},\ \bibinfo {pages} {064503}
  (\bibinfo {year} {2008})}\BibitemShut {NoStop}%
\bibitem [{\citenamefont {Cadot}\ \emph {et~al.}(2008)\citenamefont {Cadot},
  \citenamefont {Boudaoud},\ and\ \citenamefont
  {Touz{\'e}}}]{cadot2008statistics}%
  \BibitemOpen
  \bibfield  {author} {\bibinfo {author} {\bibfnamefont {O.}~\bibnamefont
  {Cadot}}, \bibinfo {author} {\bibfnamefont {A.}~\bibnamefont {Boudaoud}}, \
  and\ \bibinfo {author} {\bibfnamefont {C.}~\bibnamefont {Touz{\'e}}},\
  }\href@noop {} {\bibfield  {journal} {\bibinfo  {journal} {The European
  Physical Journal B}\ }\textbf {\bibinfo {volume} {66}},\ \bibinfo {pages}
  {399} (\bibinfo {year} {2008})}\BibitemShut {NoStop}%
\bibitem [{\citenamefont {Biferale}\ \emph {et~al.}(1998)\citenamefont
  {Biferale}, \citenamefont {Pierotti},\ and\ \citenamefont
  {Vulpiani}}]{biferale1998time}%
  \BibitemOpen
  \bibfield  {author} {\bibinfo {author} {\bibfnamefont {L.}~\bibnamefont
  {Biferale}}, \bibinfo {author} {\bibfnamefont {D.}~\bibnamefont {Pierotti}},
  \ and\ \bibinfo {author} {\bibfnamefont {A.}~\bibnamefont {Vulpiani}},\
  }\href@noop {} {\bibfield  {journal} {\bibinfo  {journal} {Journal of Physics
  A: Mathematical and General}\ }\textbf {\bibinfo {volume} {31}},\ \bibinfo
  {pages} {21} (\bibinfo {year} {1998})}\BibitemShut {NoStop}%
\bibitem [{\citenamefont {Gallavotti}\ \emph {et~al.}(2004)\citenamefont
  {Gallavotti}, \citenamefont {Rondoni},\ and\ \citenamefont
  {Segre}}]{gallavotti2004lyapunov}%
  \BibitemOpen
  \bibfield  {author} {\bibinfo {author} {\bibfnamefont {G.}~\bibnamefont
  {Gallavotti}}, \bibinfo {author} {\bibfnamefont {L.}~\bibnamefont {Rondoni}},
  \ and\ \bibinfo {author} {\bibfnamefont {E.}~\bibnamefont {Segre}},\
  }\href@noop {} {\bibfield  {journal} {\bibinfo  {journal} {Physica D:
  Nonlinear Phenomena}\ }\textbf {\bibinfo {volume} {187}},\ \bibinfo {pages}
  {338} (\bibinfo {year} {2004})}\BibitemShut {NoStop}%
\bibitem [{\citenamefont {Shang}\ \emph {et~al.}(2005)\citenamefont {Shang},
  \citenamefont {Tong},\ and\ \citenamefont {Xia}}]{shang2005test}%
  \BibitemOpen
  \bibfield  {author} {\bibinfo {author} {\bibfnamefont {X.-D.}\ \bibnamefont
  {Shang}}, \bibinfo {author} {\bibfnamefont {P.}~\bibnamefont {Tong}}, \ and\
  \bibinfo {author} {\bibfnamefont {K.-Q.}\ \bibnamefont {Xia}},\ }\href@noop
  {} {\bibfield  {journal} {\bibinfo  {journal} {Physical Review E}\ }\textbf
  {\bibinfo {volume} {72}},\ \bibinfo {pages} {015301} (\bibinfo {year}
  {2005})}\BibitemShut {NoStop}%
\bibitem [{\citenamefont {Zamponi}(2007)}]{zamponi2007possible}%
  \BibitemOpen
  \bibfield  {author} {\bibinfo {author} {\bibfnamefont {F.}~\bibnamefont
  {Zamponi}},\ }\href@noop {} {\bibfield  {journal} {\bibinfo  {journal}
  {Journal of Statistical Mechanics: Theory and Experiment}\ }\textbf {\bibinfo
  {volume} {2007}},\ \bibinfo {pages} {P02008} (\bibinfo {year}
  {2007})}\BibitemShut {NoStop}%
\bibitem [{\citenamefont {Ruelle}(2012)}]{ruelle2012hydrodynamic}%
  \BibitemOpen
  \bibfield  {author} {\bibinfo {author} {\bibfnamefont {D.~P.}\ \bibnamefont
  {Ruelle}},\ }\href@noop {} {\bibfield  {journal} {\bibinfo  {journal}
  {Proceedings of the National Academy of Sciences}\ }\textbf {\bibinfo
  {volume} {109}},\ \bibinfo {pages} {20344} (\bibinfo {year}
  {2012})}\BibitemShut {NoStop}%
\bibitem [{\citenamefont {Sekimoto}(2010)}]{sekimoto2010stochastic}%
  \BibitemOpen
  \bibfield  {author} {\bibinfo {author} {\bibfnamefont {K.}~\bibnamefont
  {Sekimoto}},\ }\href@noop {} {\emph {\bibinfo {title} {Stochastic
  energetics}}},\ Vol.\ \bibinfo {volume} {799}\ (\bibinfo  {publisher}
  {Springer},\ \bibinfo {year} {2010})\BibitemShut {NoStop}%
\bibitem [{\citenamefont {Toschi}\ and\ \citenamefont
  {Bodenschatz}(2009)}]{toschi2009lagrangian}%
  \BibitemOpen
  \bibfield  {author} {\bibinfo {author} {\bibfnamefont {F.}~\bibnamefont
  {Toschi}}\ and\ \bibinfo {author} {\bibfnamefont {E.}~\bibnamefont
  {Bodenschatz}},\ }\href@noop {} {\bibfield  {journal} {\bibinfo  {journal}
  {Annual Review of Fluid Mechanics}\ }\textbf {\bibinfo {volume} {41}},\
  \bibinfo {pages} {375} (\bibinfo {year} {2009})}\BibitemShut {NoStop}%
\bibitem [{\citenamefont {Puglisi}\ \emph {et~al.}(2005)\citenamefont
  {Puglisi}, \citenamefont {Visco}, \citenamefont {Barrat}, \citenamefont
  {Trizac},\ and\ \citenamefont {van Wijland}}]{puglisi2005fluctuations}%
  \BibitemOpen
  \bibfield  {author} {\bibinfo {author} {\bibfnamefont {A.}~\bibnamefont
  {Puglisi}}, \bibinfo {author} {\bibfnamefont {P.}~\bibnamefont {Visco}},
  \bibinfo {author} {\bibfnamefont {A.}~\bibnamefont {Barrat}}, \bibinfo
  {author} {\bibfnamefont {E.}~\bibnamefont {Trizac}}, \ and\ \bibinfo {author}
  {\bibfnamefont {F.}~\bibnamefont {van Wijland}},\ }\href@noop {} {\bibfield
  {journal} {\bibinfo  {journal} {Physical review letters}\ }\textbf {\bibinfo
  {volume} {95}},\ \bibinfo {pages} {110202} (\bibinfo {year}
  {2005})}\BibitemShut {NoStop}%
\bibitem [{\citenamefont {Puglisi}\ and\ \citenamefont
  {Villamaina}(2009)}]{puglisi2009irreversible}%
  \BibitemOpen
  \bibfield  {author} {\bibinfo {author} {\bibfnamefont {A.}~\bibnamefont
  {Puglisi}}\ and\ \bibinfo {author} {\bibfnamefont {D.}~\bibnamefont
  {Villamaina}},\ }\href@noop {} {\bibfield  {journal} {\bibinfo  {journal}
  {EPL (Europhysics Letters)}\ }\textbf {\bibinfo {volume} {88}},\ \bibinfo
  {pages} {30004} (\bibinfo {year} {2009})}\BibitemShut {NoStop}%
\bibitem [{\citenamefont {Sarracino}\ \emph {et~al.}(2010)\citenamefont
  {Sarracino}, \citenamefont {Villamaina}, \citenamefont {Gradenigo},\ and\
  \citenamefont {Puglisi}}]{sarracino2010irreversible}%
  \BibitemOpen
  \bibfield  {author} {\bibinfo {author} {\bibfnamefont {A.}~\bibnamefont
  {Sarracino}}, \bibinfo {author} {\bibfnamefont {D.}~\bibnamefont
  {Villamaina}}, \bibinfo {author} {\bibfnamefont {G.}~\bibnamefont
  {Gradenigo}}, \ and\ \bibinfo {author} {\bibfnamefont {A.}~\bibnamefont
  {Puglisi}},\ }\href@noop {} {\bibfield  {journal} {\bibinfo  {journal} {EPL
  (Europhysics Letters)}\ }\textbf {\bibinfo {volume} {92}},\ \bibinfo {pages}
  {34001} (\bibinfo {year} {2010})}\BibitemShut {NoStop}%
\bibitem [{\citenamefont {Liot}\ \emph {et~al.}(2015)\citenamefont {Liot},
  \citenamefont {Seychelles}, \citenamefont {Zonta}, \citenamefont {Chibbaro},
  \citenamefont {Coudarchet}, \citenamefont {Gasteuil}, \citenamefont {Pinton},
  \citenamefont {Salort},\ and\ \citenamefont {Chilla'}}]{liot}%
  \BibitemOpen
  \bibfield  {author} {\bibinfo {author} {\bibfnamefont {O.}~\bibnamefont
  {Liot}}, \bibinfo {author} {\bibfnamefont {F.}~\bibnamefont {Seychelles}},
  \bibinfo {author} {\bibfnamefont {F.}~\bibnamefont {Zonta}}, \bibinfo
  {author} {\bibfnamefont {S.}~\bibnamefont {Chibbaro}}, \bibinfo {author}
  {\bibfnamefont {T.}~\bibnamefont {Coudarchet}}, \bibinfo {author}
  {\bibfnamefont {Y.}~\bibnamefont {Gasteuil}}, \bibinfo {author}
  {\bibfnamefont {J.}~\bibnamefont {Pinton}}, \bibinfo {author} {\bibfnamefont
  {J.}~\bibnamefont {Salort}}, \ and\ \bibinfo {author} {\bibfnamefont
  {F.}~\bibnamefont {Chilla'}},\ }\href@noop {} {\bibfield  {journal} {\bibinfo
   {journal} {submitted to Journal of Fluid Mechanics}\ }\textbf {\bibinfo
  {volume} {-}} (\bibinfo {year} {2015})}\BibitemShut {NoStop}%
\bibitem [{\citenamefont {Zonta}\ and\ \citenamefont
  {Soldati}(2014)}]{zonta2014}%
  \BibitemOpen
  \bibfield  {author} {\bibinfo {author} {\bibfnamefont {F.}~\bibnamefont
  {Zonta}}\ and\ \bibinfo {author} {\bibfnamefont {A.}~\bibnamefont
  {Soldati}},\ }\href@noop {} {\bibfield  {journal} {\bibinfo  {journal}
  {Journal of heat transfer - Trans. ASME}\ }\textbf {\bibinfo {volume} {136}}
  (\bibinfo {year} {2014})}\BibitemShut {NoStop}%
\bibitem [{\citenamefont {Zonta}(2013)}]{zonta2013}%
  \BibitemOpen
  \bibfield  {author} {\bibinfo {author} {\bibfnamefont {F.}~\bibnamefont
  {Zonta}},\ }\href@noop {} {\bibfield  {journal} {\bibinfo  {journal}
  {International Journal of Heat and Fluid Flow}\ }\textbf {\bibinfo {volume}
  {44}},\ \bibinfo {pages} {489} (\bibinfo {year} {2013})}\BibitemShut
  {NoStop}%
\bibitem [{\citenamefont {Lebowitz}\ and\ \citenamefont
  {Spohn}(1982)}]{lebowitz1982steady}%
  \BibitemOpen
  \bibfield  {author} {\bibinfo {author} {\bibfnamefont {J.}~\bibnamefont
  {Lebowitz}}\ and\ \bibinfo {author} {\bibfnamefont {H.}~\bibnamefont
  {Spohn}},\ }\href@noop {} {\bibfield  {journal} {\bibinfo  {journal} {Journal
  of Statistical Physics}\ }\textbf {\bibinfo {volume} {29}},\ \bibinfo {pages}
  {39} (\bibinfo {year} {1982})}\BibitemShut {NoStop}%
\bibitem [{\citenamefont {Farago}(2002)}]{farago2002injected}%
  \BibitemOpen
  \bibfield  {author} {\bibinfo {author} {\bibfnamefont {J.}~\bibnamefont
  {Farago}},\ }\href@noop {} {\bibfield  {journal} {\bibinfo  {journal}
  {Journal of statistical physics}\ }\textbf {\bibinfo {volume} {107}},\
  \bibinfo {pages} {781} (\bibinfo {year} {2002})}\BibitemShut {NoStop}%
\bibitem [{\citenamefont {Gasteuil}\ \emph {et~al.}(2007)\citenamefont
  {Gasteuil}, \citenamefont {Shew}, \citenamefont {Gibert}, \citenamefont
  {Chilla'}, \citenamefont {Castaing},\ and\ \citenamefont
  {Pinton}}]{gasteuil}%
  \BibitemOpen
  \bibfield  {author} {\bibinfo {author} {\bibfnamefont {Y.}~\bibnamefont
  {Gasteuil}}, \bibinfo {author} {\bibfnamefont {W.}~\bibnamefont {Shew}},
  \bibinfo {author} {\bibfnamefont {M.}~\bibnamefont {Gibert}}, \bibinfo
  {author} {\bibfnamefont {F.}~\bibnamefont {Chilla'}}, \bibinfo {author}
  {\bibfnamefont {B.}~\bibnamefont {Castaing}}, \ and\ \bibinfo {author}
  {\bibfnamefont {J.}~\bibnamefont {Pinton}},\ }\href@noop {} {\bibfield
  {journal} {\bibinfo  {journal} {Physical Review Letters}\ }\textbf {\bibinfo
  {volume} {99}} (\bibinfo {year} {2007})}\BibitemShut {NoStop}%
\bibitem [{\citenamefont {Van~Zon}\ and\ \citenamefont
  {Cohen}(2004)}]{van2004extended}%
  \BibitemOpen
  \bibfield  {author} {\bibinfo {author} {\bibfnamefont {R.}~\bibnamefont
  {Van~Zon}}\ and\ \bibinfo {author} {\bibfnamefont {E.}~\bibnamefont
  {Cohen}},\ }\href@noop {} {\bibfield  {journal} {\bibinfo  {journal}
  {Physical Review E}\ }\textbf {\bibinfo {volume} {69}},\ \bibinfo {pages}
  {056121} (\bibinfo {year} {2004})}\BibitemShut {NoStop}%
\bibitem [{\citenamefont {Monin}\ and\ \citenamefont
  {Yaglom}(2007)}]{monin2007statistical}%
  \BibitemOpen
  \bibfield  {author} {\bibinfo {author} {\bibfnamefont {A.~S.}\ \bibnamefont
  {Monin}}\ and\ \bibinfo {author} {\bibfnamefont {A.~M.}\ \bibnamefont
  {Yaglom}},\ }\href@noop {} {\emph {\bibinfo {title} {Statistical fluid
  mechanics: mechanics of turbulence}}}\ (\bibinfo  {publisher} {Dover},\
  \bibinfo {year} {2007})\BibitemShut {NoStop}%
\bibitem [{\citenamefont {Huisman}\ \emph {et~al.}(2012)\citenamefont
  {Huisman}, \citenamefont {van Gils}, \citenamefont {Grossmann}, \citenamefont
  {Sun},\ and\ \citenamefont {Lohse}}]{lohse2}%
  \BibitemOpen
  \bibfield  {author} {\bibinfo {author} {\bibfnamefont {S.}~\bibnamefont
  {Huisman}}, \bibinfo {author} {\bibfnamefont {D.}~\bibnamefont {van Gils}},
  \bibinfo {author} {\bibfnamefont {S.}~\bibnamefont {Grossmann}}, \bibinfo
  {author} {\bibfnamefont {C.}~\bibnamefont {Sun}}, \ and\ \bibinfo {author}
  {\bibfnamefont {D.}~\bibnamefont {Lohse}},\ }\href@noop {} {\bibfield
  {journal} {\bibinfo  {journal} {Physical Review Letters}\ }\textbf {\bibinfo
  {volume} {108}} (\bibinfo {year} {2012})}\BibitemShut {NoStop}%
\bibitem [{\citenamefont {Gallavotti}(1997)}]{gallavotti1997dynamical}%
  \BibitemOpen
  \bibfield  {author} {\bibinfo {author} {\bibfnamefont {G.}~\bibnamefont
  {Gallavotti}},\ }\href@noop {} {\bibfield  {journal} {\bibinfo  {journal}
  {Physica D: Nonlinear Phenomena}\ }\textbf {\bibinfo {volume} {105}},\
  \bibinfo {pages} {163} (\bibinfo {year} {1997})}\BibitemShut {NoStop}%
\bibitem [{\citenamefont {Rondoni}\ and\ \citenamefont
  {Segre}(1999)}]{rondoni1999fluctuations}%
  \BibitemOpen
  \bibfield  {author} {\bibinfo {author} {\bibfnamefont {L.}~\bibnamefont
  {Rondoni}}\ and\ \bibinfo {author} {\bibfnamefont {E.}~\bibnamefont
  {Segre}},\ }\href@noop {} {\bibfield  {journal} {\bibinfo  {journal}
  {Nonlinearity}\ }\textbf {\bibinfo {volume} {12}},\ \bibinfo {pages} {1471}
  (\bibinfo {year} {1999})}\BibitemShut {NoStop}%
\bibitem [{\citenamefont {Rondoni}\ and\ \citenamefont
  {Morriss}(2003)}]{rondoni2003large}%
  \BibitemOpen
  \bibfield  {author} {\bibinfo {author} {\bibfnamefont {L.}~\bibnamefont
  {Rondoni}}\ and\ \bibinfo {author} {\bibfnamefont {G.~P.}\ \bibnamefont
  {Morriss}},\ }\href@noop {} {\bibfield  {journal} {\bibinfo  {journal} {Open
  Systems \& Information Dynamics}\ }\textbf {\bibinfo {volume} {10}},\
  \bibinfo {pages} {105} (\bibinfo {year} {2003})}\BibitemShut {NoStop}%
\end{thebibliography}%

\end{document}